\documentclass{article}
\usepackage{authblk}
\usepackage[table]{xcolor}
\usepackage[a4paper, margin=1in]{geometry}
\usepackage[english]{babel}
\usepackage[utf8]{inputenc}
\usepackage{array}
\usepackage{xcolor}
\usepackage{booktabs}
\usepackage{multicol}
\usepackage{longtable}
\usepackage{amsmath}
\usepackage{graphicx}
\usepackage[hyphens]{url}
\usepackage{multirow}
\providecommand{\keywords}[1]{\textbf{\textit{Index terms---}} #1}
\definecolor{Highlight}{rgb}{0.68,1,0.68}
\title{On the Effectiveness of Binary Emulation in Malware Classification}

\begin{document}
\author[1]{Vasilis Vouvoutsis}
\author[2,3]{Fran Casino}
\author[1,3]{Constantinos Patsakis}

\affil[1]{Department of Informatics, University of Piraeus, 80 Karaoli \& Dimitriou str., 18534 Piraeus, Greece}
\affil[2]{Department of Computer Engineering and Mathematics, Universitat Rovira i Virgili}
\affil[3]{Information Management Systems Institute, Athena Research Centre, Greece}

\date{}
\maketitle

\begin{abstract}
Malware authors are continuously evolving their code base to include counter-analysis methods that can significantly hinder their detection and blocking. While the execution of malware in a sandboxed environment may provide a lot of insightful feedback about what the malware actually does in a machine, anti-virtualisation and hooking evasion methods may allow malware to bypass such detection methods. The main objective of this work is to complement sandbox execution with the use of binary emulation frameworks. The core idea is to exploit the fact that binary emulation frameworks may quickly test samples quicker than a sandbox environment as they do not need to open a whole new virtual machine to execute the binary. While with this approach, we lose the granularity of the data that can be collected through a sandbox, due to scalability issues, one may need to simply determine whether a file is malicious or to which malware family it belongs. To this end, we record the API calls that are performed and use them to explore the efficacy of using them as features for binary and multiclass classification. Our extensive experiments with real-world malware illustrate that this approach is very accurate, achieving state-of-the art outcomes with a statistically robust set of classification experiments while simultaneously having a relatively low computational overhead compared to traditional sandbox approaches. In fact, we compare the binary analysis results with a commercial sandbox, and our classification outperforms it at the expense of the fine-grained results that a sandbox provides.
\end{abstract}

\section{Introduction}
Malicious software, commonly known as malware, is a piece of software that enables an attacker to fulfil her harmful intent. The continuous increase of volume and complexity of malware, coupled with the introduction of new methods, introduces many challenges in detection and blocking. The most utilised malware detection strategy is signature checking, which depends on pattern matching of known indicators of compromise (IOC) \cite{10.1145/3460120.3484759}. Whereas signature filtering is compelling for numerous sorts of malware, it is ineffectual for recognising modern malware, as, e.g. packers can easily break many such patterns. Therefore, malware authors create various malware instances from the same malware family \cite{dinh2016malware}. As a result, members of the same malware family are functionally identical, even though their binaries may greatly differ. As a result, many researchers and practitioners try to leverage machine learning to address these challenges, which seems to live up to the expectations \cite{UCCI2019123}.

One of the most widely used methods to determine the behaviour of a specific piece of software and whether it is malicious or not is to execute it inside a sandbox. Practically, we create a highly monitored and controlled environment that replicates a real user environment and execute the binary inside it to track its behaviour. Note that deviations in the execution environment may lead to radically different malware behaviour \cite{274687}. Clearly, malware analysis is something that malware authors do not want to happen; therefore, they embed several evasion methods, e.g. anti-virtualisation methods or detection of monitoring processes in the arsenal of their malware \cite{yokoyama2016sandprint,rudd2017survey,bulazel2017survey}. In fact, as shown in \cite{koutsokostas2021python}, many publicly and widely used malware sandboxes have issues with these measures, which means that they can be easily bypassed. Moreover, malware sandboxes are ``expensive'' in various ways. They operate inside a virtual machine that needs many computational and storage resources to imitate a real one. Failure to comply with these expectations will result in the detection of the analysis environment. Therefore, we need more cost-efficient ways to analyse malware. Moreover, as shown in \cite{apostolopoulos2021resurrecting} traditional anti-analysis methods could be used to unhook binaries and create many issues for the analysts as in most cases they operate with the same permissions.

Beyond sandboxes, there are several binary emulation frameworks, such as Qiling \cite{qiling}, Speakeasy \cite{speakeasy}, angr \cite{shoshitaishvili2016state}, Zelos \cite{zelos}, BitBlaze \cite{bitblaze}, and Binee \cite{Binee}, which enable analysts to emulate the execution of a binary in a host by replicating functions, system calls, and operating subsystems. The key difference here compared to sandboxing is that one can scale this better than virtual machines since binary emulation can be performed in containers that are more resource friendly. Exploiting this technology, one can understand that while in its infancy, for malware analysis, it is very promising, so its potential must be further explored. The aforementioned frameworks allow for far more detailed analysis by hooking and debugging binaries. However, this part is primarily manual and cannot scale. In this regard, the goal of this work is to investigate how binary analysis frameworks can facilitate binary (benign vs malicious scenario) as well as multiclass classification in a large and real-world dataset in a cost-effective manner. To this end, we perform binary emulation of 71,536 binaries and record their API calls to use them as features to feed machine learning models. While this does not leverage the full potentials of binary emulation, it allows us to classify malware and even classify them into families efficiently. Evidently, the results of binary emulation cannot compete in terms of granularity with the ones of traditional sandboxes. Nevertheless, we illustrate that they are enough for the classification task, using significantly less computational and storage resources.

It has to be highlighted that the construction of a malware analysis system necessitates a number of architectural considerations with far-reaching implications. Whenever a program runs in a monitoring environment, an analysis component must keep track of every element of the program's execution. The time required to reset the analysis environment to a clean state is another factor that may impact the design of an analysis technique. This is required since findings can only be compared if each sample is run in the same environment. The greater the number of samples to be examined, the greater the influence of this reset time. Clearly, containers are far more efficient than virtual machines as they are more lightweight and more cloud friendly.

Beyond scalability issues, binary emulation can also bypass some sandbox analysis limitations. For instance, when a binary is emulated, all calls are initiated by the sample under investigation. However, in a sandbox environment, since the whole operating system is being used, many calls which are recorded are initiated by processes of the operating system and have nothing to do with the analysis of the binary, introducing a lot of noise. The same applies to API hooking, where its origin might create additional noise in the analysis.

\begin{figure}[th]
    \centering
    \includegraphics[width=.6\linewidth]{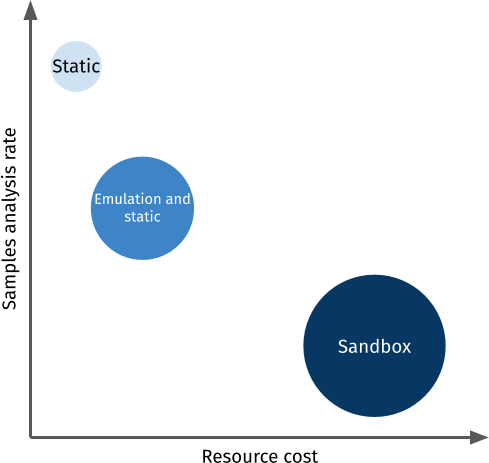}
    \caption{Positioning of the proposed approach. Sensitivity is depicted by colour, the darker the less sensitive. Insight is illustrated by size, the bigger the circle, the greater the insight. }
    \label{fig:positioning}
\end{figure}

The positioning of the proposed approach is illustrated in Figure \ref{fig:positioning}. In essence, static analysis is very efficient in terms of the time needed to analyse a sample and the resources needed for analysis. However, it is very sensitive to changes, so an adversary can easily bypass it. Moreover, only a small amount of information about what the binary can actually do can be extracted. On the contrary, dynamic analysis with a sandbox introduces high costs in resources that have to be committed and the time needed to analyse a sample is dramatically increased. However, the analysis is more robust and fine-grained, giving us good insight into what the binary actually does. In our proposal, we leverage binary emulation with static features, which significantly reduces the cost of computational resources that have to be committed while simultaneously increasing the speed that it can be performed. Moreover, since the binary is executed, we gain an insight into what the binary actually does. Due to the amount of maturity of sandboxes and the monitoring mechanisms they have, the amount of collected information by such an approach is far greater; nevertheless, it is enough to perform precise malware classification.

In this context, the main contribution of this work is to showcase in a real-world experiment the efficacy of binary emulation for malware analysis in scale. In the bulk of literature, when binary emulation is used on a large scale, the focus is on finding bugs and vulnerabilities \cite{hernandez2017firmusb,peng2018t}. Despite the continuous discussions about the use of binary emulation to analyse malware, all related work, belonging only to grey literature, refers to isolated and small scale experiments which require manual interaction. Moreover, although some of these frameworks belong to companies, none of them reports using them in production, let alone scale, but rather as a tool for their analysts to examine and interact with one sample at a time. Moreover, there is no prior work on feature extraction and machine learning methods on artefacts collected from binary emulation of binaries. Thus, to the best of our knowledge, this is the first work to test binary emulation in a large real-world dataset and exploit it to perform classification. As a result, the proposed approach requires a fraction of the actual resources and time that would be needed for sandbox analysis. Moreover, our analysis is far more robust than static and with deeper insight. In practice, we expect to use our approach to quickly filter what has to be analysed by a sandbox after an initial static analysis, significantly reducing the resources needed to analyse malware samples.

\begin{figure}[th]
    \centering
    \includegraphics[width=0.8\linewidth]{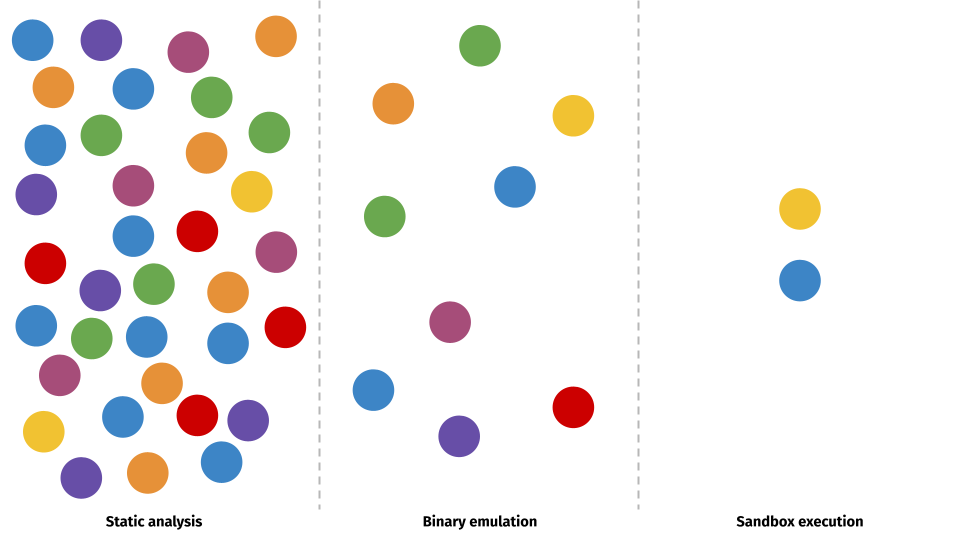}
    \caption{Filtering process of malware samples.}
    \label{fig:filtering}
\end{figure}

In Figure \ref{fig:filtering} we try to illustrate the decrease of the number of samples that have to be analysed via a sandbox. In the first step, we apply static analysis on the samples applying YARA rules, extracting strings, computing fuzzy hashes (e.g. ssdeep, TLSH) and other hashes (e.g. imphash) etc. From the extracted information, we may prune most samples, and we perform binary emulation on them to extract some dynamic features. These features allow us to prune the samples that have to be sent for further analysis as we may tell which is the family and whether it differentiates from the other samples of the family. As a result, what has to be executed in the sandbox is only a fraction of the samples of the previous step.

Finally, to prove the efficacy of our proposed pipeline, we go a step beyond merely comparing it with the current state of the art by comparing our approach to a commercial sandbox. This allows us to establish a baseline comparison in a real-world setting where both detection efficiency and resource allocation can be easily compared. While binary emulation cannot provide the fine-grained results of a sandbox analysis at this stage, the extracted features and our thorough and targeted feature engineering process allow our method to outperform a commercial sandbox in binary classification.

The rest of this work is structured as follows. In the next section, we provide an overview of the related work. In Section \ref{exp.setup}, we detail our automated pipeline for binary analysis and our data collection methodology. Then, we present our extracted dataset, discussing its composition and its fitness for training machine learning models. In Section \ref{exp.setup:class}, we present how one should train the machine learning model properly and the efficacy in the binary and multiclass setting. Next, we discuss our results and compare our work with the current state of the art. To this end, we provide a comparison with a commercial sandbox and then go through the limitations of our approach. Finally, we conclude the article summarising our contributions providing ideas for future work.

\section{Related work}

\subsection{Malware analysis}
There exist two main methods to detect and analyse malware, namely static analysis and dynamic analysis \cite{gandotra2014malware}. In static analysis, there is no execution of the file under inspection. Thus, the analysis is based solely on the contents and metadata of the file. As a result, only static features of the file under investigation are used to detect its behaviour. Typical examples can be file fragments (not necessarily continuous), imported libraries, strings, entropy, opcodes, etc., as they can be extracted from a file without executing it.

Dynamic analysis processes the actions taken by a program while it is running \cite{egele2008survey}. In this regard, a file under investigation is executed in a sandbox, a controlled and monitored environment, to determine what it does in terms of file system changes, network connections, opening processes, or performing specific system calls \cite{or2019dynamic}. Since the file is executed, the analysis does not care about the obfuscation and packing as the file will be unpacked, memory dumps can be extracted, and the proper execution path can be revealed. Beyond simply executing it, one may also try to debug the binary that has to be investigated to find out the capabilities of the file, how it performs specific tasks, and even alter its execution flow to, e.g. bypass counter-analysis measures.

These analysis methods are well-known and on most occasions, available to malware authors; thus, they try to bypass them or at least hinder them using several anti-analysis methods \cite{branco2012scientific}. Traditional measures include packers and encryption to obfuscate their contents create files with different signatures that allow them to bypass static analysis checks \cite{christodorescu2006static}. To bypass dynamic analysis, malware authors embed checks in their binary which assess the environment in which their file is executed and how it differs from a typical user environment. These checks try to determine whether the host the file is executed is monitored by, e.g. detecting user interaction (mouse movement, key pressing), checking the names of running processes for well-known analysis software, searching for environmental variables that may disclose an analysis environment, the existence of specific hooks, or even deviations from normal execution flow in terms of timing.

Of specific interest to our work is the use of API calls, as they are a very persistent characteristic that cannot be removed. The method is by no means new. One of the first works on the topic was from Forrest et al. \cite{forrest1996sense} who utilised n-grams for irregularity detection. Their tests indicate that brief sequences of system calls in operating processes provide a consistent signal of normal operation. Reddy and Pujari \cite{reddy2006n} also proposed that byte sequence and byte n-gram might be considered for feature extraction. As the number of resulting features would be exceptionally large, they utilised different strategies of feature determination and reported the use of information gain based selection strategies as the leading approach within the malware classification. Anderson et al. \cite{anderson2011graph}, introduced a unique malware detection technique based on graph analysis of dynamically gathered instruction traces of the target executable. Qiao et al. \cite{Qiao14} collected API calls through dynamic analysis using Cuckoo and enriched the features with Maleur \cite{rieck2011automatic}. Then, they transformed the features into byte sequences that were used for binary classification. Ki et al. \cite{ki2015novel} utilised DNA sequence alignment algorithms on API call sequences and found that malware follows specific patterns. Tang and Qian \cite{tang2019dynamic} also used API calls to classify malware of 9 families; however, the extracted API call sequences were converted to feature images to train a convolutional neural network to perform multiclass classification. In \cite{amer2020dynamic}, Amer and Zelinka use word embeddings to create contextual relationships between the API calls that are performed by malware and benign software to train classifiers, but also a prediction model that tries to determine whether an API call sequence is malicious or not with high accuracy.

A key point to the discussion here is how the API calls were collected. In several studies, the API calls are statically extracted \cite{aafer2013droidapiminer,amer2020dynamic} which implies that several calls are missed due to obfuscation of the binaries. To prevent such issues, researchers usually use a sandbox-like Cuckoo\footnote{\url{https://cuckoosandbox.org/}} to collect all the calls, network connections, and filesystem changes a binary makes. While most interactions tend to happen within the first two minutes \cite{kuchler2021does}, this is not a panacea. Moreover, previous malware characteristics are not necessarily good indicators as malware evolves \cite{203684}. Beyond the errors that the testing environment may have and which may impede the malware analysis due to the embedded evasive methods, malware is known to be sensitive to environmental factors and exhibit different behaviour depending on where, when, and how it is executed  \cite{balzarotti2010efficient,lindorfer2011detecting,koutsokostas2021python,274687}.

\subsection{Tools}
In the following paragraphs, we mention the core tools used that allowed the current work to be implemented. Even though those tools can be swapped effortlessly for different ones, we based our decision on the fact that they have active development, extensive communities behind them and are greatly versatile to support future work.

\subsubsection{Unicorn}
Unicorn \cite{unicorn} is a CPU emulator framework based on Qemu \cite{quynh2015unicorn}. It focuses on emulating CPU instructions that can understand emulator memory. Beyond that, Unicorn is not aware of higher-level concepts, such as dynamic libraries, system calls, I/O handling or executable formats like PE, MachO or ELF. As a result, Unicorn can only emulate raw machine instructions without Operating System (OS) context. Unicorn adds an easy-to-use API to QEMU, exposing capabilities like reading and writing memory and hooking specific locations, and memory accesses with custom callbacks \cite{maier2020basesafe}. Given a binary sample for a CPU platform, Unicorn, or QEMU for that matter, operates during run-time by executing the following steps \cite{maier2019unicorefuzz}:
\begin{enumerate}
  \item Convert a basic code block from the instruction set of the target platform to the instruction set of the host platform.
  \item Save the translated blocks in a cache.
  \item Keep a mapping from the source program counter to the destination program counter in an address lookup database.
  \item Put the translated block into action.
  \item Continue with the next found block.
\end{enumerate}
Unicorn Emulator has the ability to implement hooks for writing or reading of a specific memory \cite{jakubik2020cortex}. It can also execute the firmware instructions, and when access to the registered address occurs, the simulation will be terminated. The simulator user will then have a callback to input/output data, control interruptions, or change the simulation state. After that, the simulation will resume.

\subsubsection{Qiling.io}
Qiling \cite{qiling} is a sophisticated binary emulation framework powered by the Unicorn engine \cite{unicorn}. Qiling is intended to be a higher-level framework that uses Unicorn to simulate CPU instructions while also understanding OS. It contains executable format loaders (currently for PE, MachO, and ELF), dynamic linkers (to load and move shared libraries), syscall and IO handlers. As a result, Qiling can run executable binaries without requiring its native operating system.

Unicorn is a CPU emulator that can be scripted. Once a program makes a system call Qiling attempts to fully simulate what the host (Windows, Linux, etc.) would do. It is not easy to emulate the systems that an operating system provides. An operating system offers networking, a file system, the ability to load a binary (ELF, PE, MachO) into memory, and so on. While QEMU with complete system emulation may accomplish part of this, it lacks Unicorn's script-able control and deep analysis capabilities.

angr \cite{shoshitaishvili2016state}, for example, can be useful for focusing on certain portions of code. This is handy if you have reversed a binary sufficiently to know where to target. For example, parsers, which are generally complicated and susceptible, or locating a specific input to reach a target place (i.e. CTF challenges).
The difficulty that Qiling solves is that programs do not function in a vacuum; they are extremely dependent on the operating system on which they run. Emulating each operating system enables dynamic analysis that is not feasible with other frameworks.

\section{Experimental Setup}
\label{exp.setup}
To conduct our experiments using an automated pipeline, we opted to use two relatively mature frameworks, namely Qiling and Unicorn Engine. Beyond their maturity and features; described in the previous section, we opted for their use due to the frequent update circles that they have and the easier to process output they produce.

\begin{figure}[th]
\centering
  \includegraphics[width=\linewidth]{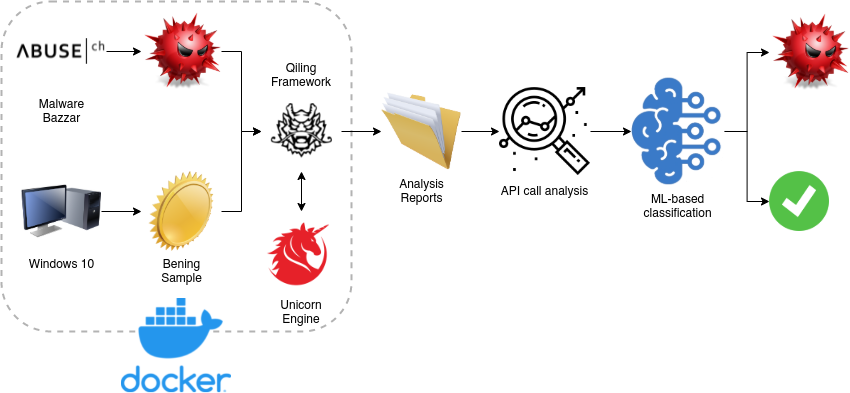}
  \caption{Overview of the developed workflow.}
  \label{fig:workflow}
\end{figure}

A generic overview of the workflow we developed is illustrated in Figure \ref{fig:workflow}. In essence, the samples are sent for execution to Qiling, which uses the Unicorn Engine to emulate their execution. The generated code runs native on the host processor, and the processor directly executes the code generated by the compiler. In our case, the instructions were processed by the Unicorn Engine and the binary was orchestrated by the Qiling binary emulation framework. Therefore, we passed each binary in Qiling, which started the binary emulation. With the help of the Unicorn engine, instructions starting from the entry point of the executable are interpreted. Finally, Qiling returns a dictionary of all API calls initiated by the analysed sample during its execution. In order for Qiling to function properly, it requires several dependencies of the emulated sample to be collected and made available to it. Hence, we performed a preliminary analysis of the samples to collect from the Windows file system all required file dependencies, e.g. DLLs. The whole execution is performed in a docker container to allow for better resource allocation and scalability. The recorded API calls are then analysed, grouped, and are used to train a machine learning model that is used for classification, both binary and multiclass. The execution of most of the samples was rather fast; however, we noticed that some of them hung for indefinite time. Since we noticed that this would happen for samples that executed more than 30sec, we introduced a timeout of one minute. Regardless of whether there was a graceful exit or a timeout, the API calls were collected. On average, the execution lasted 10.5 sec.

\subsection{Reference dataset}
In order to have a broad and realistic dataset to accurately assess our methodology, we opted to collect live and recent malware. Therefore, we downloaded from Malware Bazaar \cite{malwarebazzar}; a well-known publicly available malware repository, all available executable malware samples; denoted with mime type \texttt{x-dosexec} and file type guess \texttt{exe} in its database. From these files, we kept only the Windows PE files, non-dynamic libraries, unmanaged code executables. That is because unmanaged code executables compile straight to machine code and are directly executed by the operating system.
Thus, the total number of malicious files are 70,477.
Beyond these malware samples, we added 1,059 benign files. These files were collected from a Windows 10 installation by doing a full drive file listing and then filtering out non-PE, managed executable, that is \texttt{.dll}, \texttt{.sys}, and \texttt{.mui} files.

The resulting dataset consists of 71.536 binaries belonging to different families as shown in Table \ref{tbl:dataset}.

\begin{table}[th]
\scriptsize
\resizebox{\textwidth}{!}{
\begin{tabular}{llllllllll}
\textbf{Family} & \textbf{\#} & \textbf{Family} & \textbf{\#} & \textbf{Family} & \textbf{\#} & \textbf{Family} & \textbf{\#} & \textbf{Family} & \textbf{\#}\\
\hline
{\color[HTML]{BA55D3} \textbf{Heodo}} & 29998 & \textbf{NetSupport} & 37 & \textbf{Lazarus} & 7 & \textbf{SakulaRAT} & 3 & \textbf{VegaLocker} & 1 \\
\textbf{Unknown} & 9520 & \textbf{Riskware.Generic} & 37 & \textbf{Lu0Bot} & 7 & \textbf{MyloBot} & 3 & \textbf{JigsawLocker} & 1 \\
{\color[HTML]{BA55D3} \textbf{AgentTesla}} & 3105 & \textbf{Bunitu} & 37 & \textbf{SendSafe} & 7 & \textbf{Vidar} & 3 & \textbf{SocksBot} & 1 \\
{\color[HTML]{BA55D3} \textbf{TrickBot}} & 2739 & \textbf{Chthonic} & 36 & \textbf{FIN7} & 7 & \textbf{Adware.InstalleRex} & 3 & \textbf{GoCryptoLocker} & 1 \\
{\color[HTML]{BA55D3} \textbf{QuakBot}} & 2462 & \textbf{Meterpreter} & 36 & \textbf{Lucifer} & 7 & \textbf{BlackKingdom} & 2 & \textbf{GoldMax} & 1 \\
{\color[HTML]{BA55D3} \textbf{RaccoonStealer}} & 1930 & \textbf{VTFlooder} & 36 & \textbf{Adware.Eorezo} & 7 & \textbf{Suncrypt} & 2 & \textbf{Snojan} & 1 \\
{\color[HTML]{BA55D3} \textbf{Loki}} & 1907 & \textbf{GCleaner} & 35 & \textbf{StrongPity} & 7 & \textbf{DarktrackRAT} & 2 & \textbf{Gootkit} & 1 \\
{\color[HTML]{BA55D3} \textbf{FormBook}} & 1854 & \textbf{DarkVNC} & 35 & \textbf{Citadel} & 6 & \textbf{Djvu} & 2 & \textbf{Rasftuby} & 1 \\
{\color[HTML]{BA55D3} \textbf{RedLineStealer}} & 1650 & \textbf{Socelars} & 33 & \textbf{Loda} & 6 & \textbf{AvosLocker} & 2 & \textbf{Siplog} & 1 \\
{\color[HTML]{BA55D3} \textbf{RemcosRAT}} & 1353 & \textbf{PandaZeuS} & 33 & \textbf{RevCodeRAT} & 6 & \textbf{RTM} & 2 & \textbf{H1N1} & 1 \\
{\color[HTML]{BA55D3} \textbf{AveMariaRAT}} & 1232 & \textbf{RemoteManipulator} & 31 & \textbf{Ranzy} & 6 & \textbf{Cuba} & 2 & \textbf{Hamweq} & 1 \\
{\color[HTML]{BA55D3} \textbf{ArkeiStealer}} & 850 & \textbf{TaurusStealer} & 30 & \textbf{Matrix} & 6 & \textbf{FuxSocy} & 2 & \textbf{Shiotob} & 1 \\
{\color[HTML]{BA55D3} \textbf{NanoCore}} & 819 & \textbf{Adware.ExtenBro} & 29 & \textbf{TVRat} & 6 & \textbf{Spambot.Kelihos} & 2 & \textbf{ShikataGaNai} & 1 \\
{\color[HTML]{BA55D3} \textbf{Gozi}} & 742 & \textbf{DarkSide} & 27 & \textbf{Bancteian} & 6 & \textbf{Adwind} & 2 & \textbf{Reconyc} & 1 \\
{\color[HTML]{BA55D3} \textbf{CobaltStrike}} & 668 & \textbf{KPOTStealer} & 27 & \textbf{Turla} & 5 & \textbf{RanzyLocker} & 2 & \textbf{Jackpot} & 1 \\
{\color[HTML]{BA55D3} \textbf{Smoke Loader}} & 436 & \textbf{Babuk} & 27 & \textbf{KINS} & 5 & \textbf{Arechclient2} & 2 & \textbf{Renamer} & 1 \\
{\color[HTML]{BA55D3} \textbf{NetWire}} & 387 & \textbf{YoungLotus} & 27 & \textbf{Blackmoon} & 5 & \textbf{ZeusSphinx} & 2 & \textbf{OzoneRAT} & 1 \\
{\color[HTML]{BA55D3} \textbf{ModiLoader}} & 383 & \textbf{Cerber} & 26 & \textbf{Worm.Dorkbot} & 5 & \textbf{nccTrojan} & 2 & \textbf{Retefe} & 1 \\
{\color[HTML]{BA55D3} \textbf{BazaLoader}} & 364 & \textbf{TriumphLoader} & 24 & \textbf{Adware.InstallCore} & 5 & \textbf{Adware.Adload} & 2 & \textbf{Milum} & 1 \\
{\color[HTML]{BA55D3} \textbf{njrat}} & 355 & \textbf{Andromeda} & 22 & \textbf{Zegost} & 5 & \textbf{M00nD3v} & 2 & \textbf{Micropsia} & 1 \\
{\color[HTML]{BA55D3} \textbf{MassLogger}} & 341 & \textbf{RevengeRAT} & 20 & \textbf{SchoolBoy} & 5 & \textbf{Ryuk} & 2 & \textbf{Rustyloader} & 1 \\
{\color[HTML]{BA55D3} \textbf{AZORult}} & 328 & \textbf{BazarCall} & 20 & \textbf{LockFile} & 5 & \textbf{Adware.DownloadAdmin} & 2 & \textbf{Mercurial} & 1 \\
{\color[HTML]{BA55D3} \textbf{CryptBot}} & 326 & \textbf{Hancitor} & 20 & \textbf{DoejoCrypt} & 5 & \textbf{Xtrat} & 2 & \textbf{LodaRAT} & 1 \\
{\color[HTML]{BA55D3} \textbf{DanaBot}} & 322 & \textbf{Troldesh} & 20 & \textbf{Locky} & 5 & \textbf{Mimikatz} & 2 & \textbf{Scarab} & 1 \\
{\color[HTML]{00A933} \textbf{IcedID}} & 287 & \textbf{Zeppelin} & 19 & \textbf{StormKitty} & 5 & \textbf{Qulab} & 2 & \textbf{LolKek} & 1 \\
{\color[HTML]{00A933} \textbf{Stop}} & 278 & \textbf{CyberGate} & 18 & \textbf{FinderBot} & 5 & \textbf{Jigsaw} & 2 & \textbf{MSILStealer} & 1 \\
{\color[HTML]{00A933} \textbf{SnakeKeylogger}} & 230 & \textbf{LockBit} & 18 & \textbf{FlawedAmmyy} & 5 & \textbf{Adware.LoadMoney} & 2 & \textbf{Sazoora} & 1 \\
{\color[HTML]{00A933} \textbf{HawkEye}} & 230 & \textbf{Osiris} & 18 & \textbf{RagnarLocker} & 5 & \textbf{Adware.PushWare} & 2 & \textbf{Maener} & 1 \\
{\color[HTML]{00A933} \textbf{Amadey}} & 201 & \textbf{Mespinoza} & 17 & \textbf{Maze} & 5 & \textbf{IRCbot} & 2 & \textbf{MaktubLocker} & 1 \\
{\color[HTML]{00A933} \textbf{Pony}} & 194 & \textbf{MedusaLocker} & 17 & \textbf{PhoenixKeylogger} & 5 & \textbf{Hive} & 2 & \textbf{Maoloa} & 1 \\
{\color[HTML]{00A933} \textbf{PlugX}} & 167 & \textbf{Cutwail} & 16 & \textbf{Worm.Ramnit} & 4 & \textbf{Urelas} & 2 & \textbf{GlobeImposter} & 1 \\
{\color[HTML]{00A933} \textbf{Adware.Generic}} & 161 & \textbf{Tofsee} & 16 & \textbf{Vjw0rm} & 4 & \textbf{ShipUp} & 2 & \textbf{Gelsemium} & 1 \\
{\color[HTML]{00A933} \textbf{Orbus}} & 154 & \textbf{GoldenSpy} & 16 & \textbf{Nemty} & 4 & \textbf{Grandoreiro} & 2 & \textbf{Solaso} & 1 \\
{\color[HTML]{00A933} \textbf{GandCrab}} & 154 & \textbf{CoinMiner.XMRig} & 15 & \textbf{IAmTheKing} & 4 & \textbf{Ardamax} & 2 & \textbf{Foudre} & 1 \\
{\color[HTML]{00A933} \textbf{QuasarRAT}} & 152 & \textbf{Sage} & 15 & \textbf{WannaCry} & 4 & \textbf{DemonWare} & 2 & \textbf{CobianRAT} & 1 \\
{\color[HTML]{00A933} \textbf{ZeuS}} & 143 & \textbf{Nitol} & 15 & \textbf{StealthWorker} & 4 & \textbf{BlueBot} & 1 & \textbf{CoderWare} & 1 \\
{\color[HTML]{00A933} \textbf{BitRAT}} & 140 & \textbf{ISRStealer} & 15 & \textbf{MyDoom} & 4 & \textbf{BlackNET} & 1 & \textbf{VHDLocker} & 1 \\
{\color[HTML]{00A933} \textbf{DiamondFox}} & 134 & \textbf{Avaddon} & 15 & \textbf{HelloKitty} & 4 & \textbf{BlackRAT} & 1 & \textbf{VHD} & 1 \\
{\color[HTML]{00A933} \textbf{FickerStealer}} & 133 & \textbf{WastedLocker} & 15 & \textbf{LegionLoader} & 4 & \textbf{Blackbone} & 1 & \textbf{CrimsonRAT} & 1 \\
{\color[HTML]{00A933} \textbf{Dridex}} & 129 & \textbf{ObliqueRAT} & 14 & \textbf{TinyNuke} & 4 & \textbf{BlackRose} & 1 & \textbf{Crypt888} & 1 \\
{\color[HTML]{00A933} \textbf{OskiStealer}} & 121 & \textbf{Neshta} & 14 & \textbf{Zatoxp} & 4 & \textbf{SaintBot} & 1 & \textbf{Upatre} & 1 \\
{\color[HTML]{00A933} \textbf{AsyncRAT}} & 120 & \textbf{a310Logger} & 14 & \textbf{PandaStealer} & 4 & \textbf{BillGates} & 1 & \textbf{UniWinniCrypt} & 1 \\
{\color[HTML]{00A933} \textbf{Sodinokibi}} & 113 & \textbf{PredatorStealer} & 14 & \textbf{Ostap} & 4 & \textbf{Adware.Techsnab} & 1 & \textbf{RansomEXX} & 1 \\
\textbf{Glupteba} & 97 & \textbf{Macoute} & 13 & \textbf{Dorv} & 4 & \textbf{modi} & 1 & \textbf{TorrentLocker} & 1 \\
\textbf{DarkComet} & 93 & \textbf{Downloader.Upatre} & 13 & \textbf{LuminosityLink} & 4 & \textbf{Zeoticus} & 1 & \textbf{TeslaCrypt} & 1 \\
\textbf{DCRat} & 92 & \textbf{Makop} & 13 & \textbf{Ramnit} & 4 & \textbf{Adware.CloudScout} & 1 & \textbf{Ransomware} & 1 \\
\textbf{ParallaxRAT} & 88 & \textbf{LimeRAT} & 12 & \textbf{Xorist} & 4 & \textbf{Adware.Duote} & 1 & \textbf{TA505} & 1 \\
\textbf{ServHelper} & 87 & \textbf{Neutrino} & 12 & \textbf{SysVenFak} & 4 & \textbf{Pykspa} & 1 & \textbf{Ransomware.Nemty} & 1 \\
\textbf{Adware.Breitschopp} & 87 & \textbf{OrcusRAT} & 11 & \textbf{Anyplace} & 4 & \textbf{Pojie} & 1 & \textbf{OOO CM} & 1 \\
\textbf{CoinMiner} & 85 & \textbf{WSHRAT} & 11 & \textbf{BlackShades} & 4 & \textbf{Adware.FlyStudio} & 1 & \textbf{NukeSped} & 1 \\
\textbf{Matiex} & 83 & \textbf{Kovter} & 10 & \textbf{Bandook} & 4 & \textbf{Adware.InstallMonster} & 1 & \textbf{EduRansom} & 1 \\
\textbf{SystemBC} & 77 & \textbf{BlackMatter} & 10 & \textbf{Deathransom} & 3 & \textbf{Adware.Qjwmonkey} & 1 & \textbf{Encrpt3d} & 1 \\
\textbf{Drolnux} & 67 & \textbf{Shifu} & 10 & \textbf{ShurkStealer} & 3 & \textbf{XpertRAT} & 1 & \textbf{Erica} & 1 \\
\textbf{ZLoader} & 66 & \textbf{CryLock} & 10 & \textbf{Floxif} & 3 & \textbf{BigLock} & 1 & \textbf{Exorcist} & 1 \\
\textbf{TeamBot} & 64 & \textbf{Snatch} & 9 & \textbf{Worm.Virut} & 3 & \textbf{AnyDesk} & 1 & \textbf{Ransomware.Petya} & 1 \\
\textbf{Adware.FileTour} & 58 & \textbf{Expiro} & 8 & \textbf{FONIX} & 3 & \textbf{AresRAT} & 1 & \textbf{NexusStealer} & 1 \\
\textbf{GuLoader} & 58 & \textbf{Gh0stRAT} & 8 & \textbf{Bancos} & 3 & \textbf{Wintenzz} & 1 & \textbf{Netsky} & 1 \\
\textbf{BuerLoader} & 57 & \textbf{MountLocker} & 8 & \textbf{Pitou} & 3 & \textbf{Winlock} & 1 & \textbf{PyXie} & 1 \\
\textbf{Tinba} & 53 & \textbf{Balaclava} & 8 & \textbf{Adware.Koutodoor} & 3 & \textbf{WellMess} & 1 & \textbf{FlawedAmmyyRAT} & 1 \\
\textbf{NetWalker} & 52 & \textbf{Phobos} & 8 & \textbf{BozokRAT} & 3 & \textbf{BabylonRAT} & 1 & \textbf{Fonix} & 1 \\
\textbf{Conti} & 49 & \textbf{VMZeuS} & 8 & \textbf{VirLock} & 3 & \textbf{RDAT} & 1 & \textbf{SpyEye} & 1 \\
\textbf{Dharma} & 46 & \textbf{Redosdru} & 7 & \textbf{ClipBanker} & 3 & \textbf{Beastdoor} & 1 & \textbf{PredatorTheThief} & 1 \\
\textbf{Neurevt} & 45 & \textbf{1xxbot} & 7 & \textbf{Warezov} & 3 & \textbf{Bentley} & 1 & \textbf{} &  \\
\textbf{404Keylogger} & 43 & \textbf{ImminentRAT} & 7 & \textbf{BreakWin} & 3 & \textbf{Carbanak} & 1 & \textbf{Benign} & 1059 \\
\textbf{Phorpiex} & 39 & \textbf{Nefilim} & 7 & \textbf{Clop} & 3 & \textbf{TinyLoader} & 1 & \textbf{} & \\
\end{tabular}
 }
\caption{Composition of our Dataset. The families highlighted in green and purple were used in the multiclass classification experiments in Section \ref{exp.setup:mul:a}. Only the families highlighted in purple were used in the experiments of Section \ref{exp.setup:mul:b}. }
\label{tbl:dataset}
\end{table}

\subsection{Feature extraction}
The resulting reports for each sample from Qiling were collected, and the API calls were extracted. Further to the API call extraction, we implemented a unification step to reduce the sparseness of the dataset and group similar artefacts together. This part is essential as there are too many different API calls which, if used as is, would end up to a much more sparse dataset than the one obtained (which is already sparse). Note that sparse datasets entail further challenges when trying to derive statistical properties and correlations of a dataset, affecting the prediction accuracy of machine learning methods in multiple contexts \cite{kuss2002global,9032138,casino2019privacy,li2016sparseness}.

To this end, the unification involved the following steps:
\begin{itemize}
  \item \textbf{API counter:} For each generated report, we counted every reported API call.
  \item \textbf{ASCII/Unicode Win32 API merge:} If any of the API's arguments takes a string, Win32 provides two versions of the API. These are the API's ASCII and Unicode versions, which produce the letters A and W, respectively. The ASCII version of the API accepts ASCII strings, whereas the Unicode version allows Unicode wide character strings \cite{mohanta2020windows,msndstrings}.
  \item \textbf{Extended Win32 API merge:} Some Win32 APIs have a more extensive version. An API's expanded version is denoted by the suffix `Ex'. The distinction between a non-extended and extended API is that the extended version may take more parameters/arguments and may also provide extra functionality \cite{mohanta2020windows}. Many compromises had to be introduced to facilitate both the old 16-bit Windows API and the new 32-bit Windows API. There are numerous API functions that are available to both APIs; hence, one of them had to change, and for backward compatibility, it is in almost all cases the 32-bit version.
  \item \textbf{C runtime library:} The Microsoft runtime library provides routines for programming the Microsoft Windows operating system. These routines automate many common programming tasks that are not provided by the C and C++ languages \cite{msndcrt}. Many of these routines have an identical implementation with differences in the routine's name.
  \item \textbf{Interesting APIs}. We chose to extract the arguments of the called Win32 API and include them in our analysis. The APIs used are:
  \begin{itemize}
    \item \texttt{LoadLibrary*} function (\texttt{libloaderapi.h}). Loads the specified module into the address space of the calling process. Other modules may be loaded as a result of the given module. Malware developers frequently opt not to employ load-time linking when calling Windows APIs. Instead, they choose to disguise such calls by dynamically loading and resolving API references.
    \item \texttt{GetModuleHandle*} function (\texttt{libloaderapi.h}). Retrieves a module handle for the specified module. The calling process must have loaded the module. Some modules are loaded by default into all non-native-system processes. Such modules are \texttt{ntdll.dll} and \texttt{kernel32.dll}, therefore you do not need to call \texttt{LoadLibrary}/\texttt{FreeLibrary} on these and can instead just call \texttt{GetModuleHandle}.
    \item \texttt{GetProcAddress} function (\texttt{libloaderapi.h}). Retrieves the address of an exported function or variable from the specified dynamic-link library (DLL). Most malware researchers are aware of how APIs can be resolved dynamically via a call to \texttt{GetProcAddress}, so they can trace down all calls to that function and check the second parameter, or \texttt{lpProcName}, for the presence of high-risk APIs (or those that can download and run the malware's actual payload.) This usually helps them to narrow down and eliminate API calls that do not introduce a considerable security risk.
  \end{itemize}
  \item \textbf{Name mangling:} Name mangling is the process of encoding function and variable names into unique names so that linkers can differentiate common names in the language. Some names use a mangling format similar to Visual C/C++ \cite{agner2014calling}. In that scenario, our strategy is to use the name's strings to produce a signature generated of the name based on Bonfante et al., \cite{bonfante2016function} but more extended in order to not lose most of the included information.
\end{itemize}

\section{Classification Experiments}
\label{exp.setup:class}
In what follows, we describe our experimental setup and the tests performed to showcase the efficacy of our method. The first set of experiments focused on the enhancement of the dataset via a feature selection optimisation method, namely Boruta \cite{kursa2010feature}, and the binary classification. Next, we leverage the Random Forest model to perform a multiclass classification of malware samples. Note that, in addition to the classification models used in our experiments, we studied models such as Support Vector Classifier (SVC) and Gaussian Process Classifier (GPC), yet further experiments were discarded due to the poor scalability of such models when applied to high-dimensional datasets, requiring considerable memory resources and/or time to obtain similar or even slightly worse outcomes than the rest of the models.

\subsection{Feature selection and binary classification}
\label{exp.setup:bin}

We assess the power of our proposed features to differentiate between malicious and benign samples (i.e., binary classification). We selected a set of machine learning methods to leverage a binary classification task. More concretely, we used Random Forest, a non-parametric ensemble classifier, XGBoost, which implements gradient boosted decision trees, and a k-nearest neighbours classifier (KNN).

Previously to tuning the hyperparameters of each model, and after observing the sparseness of the dataset (the 98.78\% of the dataset's possible values are empty), we applied a feature selection process to it intending to reduce the dimensionality of the database. The latter allows us to observe which features are the most relevant and increases the efficiency of the classification task.
We selected Boruta \cite{kursa2010feature} as the methodology to perform feature selection optimisation. Boruta is an iterative method that performs a top-down search for relevant features by comparing their importance with importance achievable at random (i.e. using permutations of different feature values, namely shadow copies as declared by the authors), and discarding the ones that are irrelevant for the classification task. Since Boruta is applied to ensemble classifiers such as Random Forest, we performed our optimisations using such a classification model. We applied Boruta with the parameters shown in Table \ref{tab:modelgrid}. The outcomes of the procedure allowed us to discard 2278 features. Therefore, the binary classification experiments will be performed over a new dataset with 239 features instead of 2536 from the original dataset.

\begin{table*}[!ht]
\centering
\caption{Hyperparameters of the feature selection model and the classification methods.}
\label{tab:modelgrid}
\begin{tabular}{>{\bfseries}lp{3.25in}}
\toprule
\textbf{Model} & \textbf{Best configuration} \\
\midrule
Boruta  & learning\_rate= 0.05, max\_iter= 50, max\_depth= 5,  perc= 90 \\
\hline
Random Forest & n\_estimators= 100, max\_depth= 20\\
XGBoost & learning\_rate= 0.02, max\_depth= 10, subsample= 0.8\\
KNN & n\_neighbors= 5, weights= distance \\
\bottomrule
\end{tabular}

\end{table*}

The hyperparameters of each classification model were tuned with a grid search to maximise classification performance in the task of distinguishing between benign and malicious samples. Table \ref{tab:modelgrid} summarises the configuration parameters that achieved the highest performance. It is important to note that hyperparameter optimisation is a crucial task closely tied with the robustness of the training procedure. In this regard, we could observe that the max\_depth parameter was very relevant when maximising the efficacy of the tree-based models, increasing the possible combinations of features while exploring data interrelationships. In the case of the KNN model, using five neighbours and a weighted distance metric improved the outcomes, probably leveraging the inter-family similarities. Thus, correctly understanding the models' parameters is crucial when combined with datasets exhibiting specific features and characteristics such as our sparse, high-dimensional dataset.

The training procedure and the granularity of the metrics used are also critical. Due to the reliability of the models, the outcomes reported during all training procedures obtained F1-scores above 98\%. However, and due to the unbalanced nature of the dataset, reporting outstanding outcomes for the malicious class could cover a very low performance in the case of benign class. The latter was happening in all cases when the default parameters were used during the hyperparameter optimisation phase. In such cases, the classification outcomes of the benign class were very low or close to 0. Therefore, to produce robust methods that can classify with high reliability all classes, the classification reports should provide us with granular information both during the process of hyperparameter optimisation and during training. Finally, we employed standard 10-fold cross-validation and repeated such an experiment three times to get a roughly unbiased estimate of the performance of our trained predictive models.

To ensure the replicability of our experiments, we used the popular platform Google Colab \footnote{https://research.google.com/colaboratory/} in its free version (2x 2.2GHz CPU and 12GB of RAM), while we utilised the implementations of the \texttt{scikit-learn}\footnote{\url{https://scikit-learn.org}} library. Note that such specifications establish a bottom-line that low-mid performance desktop computers can easily outperform. We evaluate the efficacy of the trained classifiers using the standard classification metrics of precision, recall, and $F_1$ score.

As already discussed, since our dataset has 1059 benign and 70,477 malicious samples, we report the classification outcomes per class in Table \ref{tab:detailclassification}. As it can be observed, such outcomes highlight the robustness of our methodology in identifying both benign and malicious samples. Note that in unbalanced datasets, aggregated outcomes may hide critical errors in underrepresented classes. As observed in Table \ref{tab:detailclassification}, our method identified the malicious samples with outstanding reliability on average regardless of the classification model used (i.e. the outcomes of RF and KNN were almost identical), yet several benign samples were considered malicious due to the use of features that in some cases overlapped with these used by malicious ones. However, the latter guarantees the security level that we aimed in our system (i.e., misclassifying a malicious sample would incur further security issues than misclassifying a benign sample).

\begin{table*}[]
\scriptsize
\setlength{\tabcolsep}{1.2pt} 
\renewcommand{\arraystretch}{1.5} 
\rowcolors{1}{white}{gray!25}
\caption{Average outcomes and their corresponding standard deviation $\sigma$. The best F1-score outcomes are highlighted in green.}
\label{tab:detailclassification}
\resizebox{\columnwidth}{!}{%
\begin{tabular}{l|cccccc|cccccc}
& \multicolumn{6}{c|}{\textbf{Benign}} & \multicolumn{6}{c}{\textbf{Malicious}}  \\
\hline
\textbf{Method}  & \multicolumn{2}{c}{\textbf{precision}} & \multicolumn{2}{c}{\textbf{recall}} & \multicolumn{2}{c|}{\textbf{f1}} & \multicolumn{2}{c}{\textbf{precision}} & \multicolumn{2}{c}{\textbf{recall}} & \multicolumn{2}{c}{\textbf{f1}} \\
 & \textbf{Avg.} & \textbf{$\sigma$} & \textbf{Avg.} & \textbf{$\sigma$} & \textbf{Avg.} & \textbf{$\sigma$} & \textbf{Avg.} & \textbf{$\sigma$} & \textbf{Avg.} & \textbf{$\sigma$} & \textbf{Avg.} & \textbf{$\sigma$} \\
\textbf{RF} & 0.962 & 0.020 & 0.875 & 0.001 &  0.916 & 0.046 & 0.998 & \textless{}0.001 & 0.999 & 0.025 &0.999 & \textless{}0.001 \\
\textbf{XGBoost} & 0.949 & 0.021 & 0.862 & 0.037 & 0.903 & 0.024 & 0.998 & 0.001 & 0.999 & \textless{}0.001 & 0.999 & \textless{}0.001  \\
\textbf{KNN} & 0.944 & 0.023 & 0.891 & 0.031 &\cellcolor{Highlight} 0.916 & 0.020 & 0.998 & \textless{}0.001 & 0.999 & \textless{}0.001 & \cellcolor{Highlight} 0.999 & \textless{}0.001 \\
\hline
\hline
&\multicolumn{6}{c|}{\textbf{Micro average}} & \multicolumn{6}{c}{\textbf{Macro average}} \\
\textbf{Method}  & \multicolumn{2}{c}{\textbf{precision}} & \multicolumn{2}{c}{\textbf{recall}} & \multicolumn{2}{c|}{\textbf{f1}} & \multicolumn{2}{c}{\textbf{precision}} & \multicolumn{2}{c}{\textbf{recall}} & \multicolumn{2}{c}{\textbf{f1}} \\
 & \textbf{Avg.} & \textbf{$\sigma$} & \textbf{Avg.} & \textbf{$\sigma$} & \textbf{Avg.} & \textbf{$\sigma$} & \textbf{Avg.} & \textbf{$\sigma$} & \textbf{Avg.} & \textbf{$\sigma$} & \textbf{Avg.} & \textbf{$\sigma$} \\
\textbf{RF} & 0.998 & 0.001 & 0.998 & 0.001 & 0.998 & 0.001 & 0.980 & 0.010 & 0.937 & 0.023 & 0.957 & 0.013 \\
\textbf{XGBoost} & 0.997 & 0.001 & 0.997 & 0.001 & 0.997 & 0.001 & 0.974 & 0.011 & 0.931 & 0.018 & 0.951 & 0.012 \\
\textbf{KNN} & 0.998 & \textless{}0.001 & 0.998 & \textless{}0.001 & 0.998 \cellcolor{Highlight}& 0.001 & 0.971 & 0.012 & 0.945 & 0.016 &\cellcolor{Highlight} 0.958 & 0.010\\
\hline
\end{tabular}
}
\end{table*}

 In addition to the previous experiment, to further showcase the robustness of our methodology, we performed another binary classification experiment. While using all the benign samples, we randomly sampled malicious ones to obtain a dataset with a 1:1 ratio between benign and malicious. We repeated this procedure to create 100 different datasets (i.e. due to the high amount of malicious samples) and performed a binary classification applying 10-fold cross-validation in each of such datasets. The average outcomes of such an experiment are reported in Table \ref{tab:binarysampling}. The outcomes are more smooth, especially in terms of recall in the benign samples. Note that the higher the number of malicious samples, the more complex feature overlapping the classifier needs to distinguish. Thus, reducing such unbalance benefited the outcomes of the benign class while slightly decreasing the performance of the malicious ones due to the lower amount of training data available. A slight increase in the average $\sigma$ values is observed in the case of the malicious samples, which is understandable due to the use of disparate sets in each sampled dataset. Finally, note that the aim of this experiment is to reflect that the features and methods used are statistically sound regardless of the unbalanced nature of our dataset.

\begin{table}[th]
\centering
\scriptsize
\setlength{\tabcolsep}{1.2pt} 
\renewcommand{\arraystretch}{1.5} 
\rowcolors{2}{gray!25}{white}
\caption{Average outcomes and their corresponding standard deviation $\sigma$ in the binary classification with a 1:1 ratio between benign and malicious samples. The best F1-score outcomes were highlighted in green.}
\begin{tabular}{l|cccccc|cccccc}
& \multicolumn{6}{c|}{\textbf{Benign}} & \multicolumn{6}{c}{\textbf{Malicious}}  \\
\cline{2-13}
\textbf{Method}  & \multicolumn{2}{c}{\textbf{precision}} & \multicolumn{2}{c}{\textbf{recall}} & \multicolumn{2}{c|}{\textbf{f1}} & \multicolumn{2}{c}{\textbf{precision}} & \multicolumn{2}{c}{\textbf{recall}} & \multicolumn{2}{c}{\textbf{f1}}  \\
\hline
 & \textbf{Avg.} & \textbf{$\sigma$} & \textbf{Avg.} & \textbf{$\sigma$} & \textbf{Avg.} & \textbf{$\sigma$} & \textbf{Avg.} & \textbf{$\sigma$} & \textbf{Avg.} & \textbf{$\sigma$} & \textbf{Avg.} & \textbf{$\sigma$} \\
\textbf{RF} & 0.963 & 0.020 & 0.977 & 0.015 & \cellcolor{Highlight}0.970 & 0.012 & 0.977 & 0.014 & 0.962 & 0.021 &\cellcolor{Highlight} 0.969 & 0.013  \\
\textbf{XGBoost} & 0.967 & 0.017 & 0.957 & 0.020 & 0.962 & 0.013 & 0.958 & 0.019 & 0.967 & 0.018 & 0.962 & 0.013  \\
\textbf{KNN} & 0.970 & 0.017 & 0.968 & 0.017 & 0.969 & 0.012 & 0.968 & 0.016 & 0.970 & 0.018 & 0.969 & 0.012 \\
\end{tabular}
\newline
\vspace*{0.5 cm}
\newline
\rowcolors{1}{white}{gray!25}
\begin{tabular}{l|cccccc}
&  \multicolumn{6}{c}{\textbf{Total}}  \\
\hline
\textbf{Method}  & \multicolumn{2}{c}{\textbf{precision}} & \multicolumn{2}{c}{\textbf{recall}} & \multicolumn{2}{c}{\textbf{f1}} \\
\hline
 & \textbf{Avg.} & \textbf{$\sigma$} & \textbf{Avg.} & \textbf{$\sigma$} & \textbf{Avg.} & \textbf{$\sigma$}  \\
\textbf{RF}  & 0.970 & 0.017 & 0.969 & 0.018 &\cellcolor{Highlight} 0.969 & 0.013 \\
\textbf{XGBoost} & 0.962 & 0.018 & 0.962 & 0.019 & 0.962 & 0.013  \\
\textbf{KNN} & 0.969 & 0.016 & 0.969 & 0.017 & 0.969 & 0.012 \\
\end{tabular}
\label{tab:binarysampling}
\end{table}

\subsection{Multiclass classification}
\label{exp.setup:mul}

In this experiment, we aim to classify the families in a multiclass classification setting. Due to the unbalanced nature of the dataset and due to the inability of classifiers to be trained in cases where there are not enough samples \cite{casino2021intercepting}, we selected two subsets of well-represented families in our dataset to perform a multiclass classification, as described in the following sections. Further to the dynamic features, in our features, we added one static feature, the imphash of each sample. Note that since each sample is a PE executable, one can extract the imported libraries and compute the imphash.

Practically, imphash is the MD5 hash of the ordered list of lower case function names and the DLL names of a PE file. It has been used for malware clustering in many use cases  \cite{moran2013supply,mandiant}. Since the hash is computed over the imported libraries and functions, it is partially correlated with the extracted features as some functions might not have been triggered in the dynamic analysis.

\subsubsection{Dataset D100}
\label{exp.setup:mul:a}

For this experiment, we set up an eligibility threshold of 100 samples to create a new dataset, namely D100. Thus, the 42 families that fulfilled such conditions were selected, as highlighted in Table \ref{tbl:dataset}. Next, to avoid biased outcomes derived from over-represented classes, we created a balanced dataset by randomly sampling 100 elements of each family. We repeated this procedure (i.e. creating a new dataset each time) 100 times, and for each dataset, we applied 10-fold cross-validation using Random Forest classifier to leverage the multiclass classification (i.e. KNN and XGBoost performed slightly worse; thus, we used only RF for the sake of clarity). The obtained outcomes are depicted in Table \ref{tab:outcomesmulticlassA}. Note that we used the original dataset with all the features (i.e. 2536 features) for this experiment to ensure that the classifier was able to observe all the possible feature overlapping between malware families. As it can be observed, from the 42 families, 10 were identified with F1-score close to or above 0.8 (i.e. Orbus, GandCrab, Plugx, TrickBot, Heodo, BazaLoader, Sodinokibi, IcedID, QuakBot, Gozi), with Orbus, GrandCrab and Plugx reporting F1-score values above 0.95. Most of the families that obtained very low F1-scores on average obtained a substantially higher precision than recall, a common problem of multiclass classification, especially when the classes are not well-separated \cite{silva2017improving}.

\begin{table}[th]
\centering
\rowcolors{2}{gray!25}{white}
\small
\begin{tabular}{lcc|cc|cc}

\multicolumn{1}{c}{\multirow{1}{*}{\textbf{Family}}} & \multicolumn{2}{c}{\textbf{Precision}} & \multicolumn{2}{c}{\textbf{Recall}} & \multicolumn{2}{c}{\textbf{F1}} \\
\hline
\multicolumn{1}{c}{} & \multicolumn{1}{c}{Average} & \multicolumn{1}{c}{$\sigma$} & \multicolumn{1}{c}{Average} & \multicolumn{1}{c}{$\sigma$} & \multicolumn{1}{c}{Average} & \multicolumn{1}{c}{$\sigma$} \\
\hline
\textbf{Orbus} & 0.982 & 0.013 & 1.000 & 0.002 & 0.991 & 0.007 \\
\textbf{GandCrab} & 0.964 & 0.017 & 0.985 & 0.007 & 0.974 & 0.010 \\
\textbf{PlugX} & 0.986 & 0.012 & 0.952 & 0.013 & 0.969 & 0.010 \\
\textbf{TrickBot} & 0.889 & 0.036 & 0.883 & 0.032 & 0.886 & 0.028 \\
\textbf{Heodo} & 0.834 & 0.032 & 0.890 & 0.031 & 0.861 & 0.025 \\
\textbf{BazaLoader} & 0.808 & 0.032 & 0.911 & 0.031 & 0.856 & 0.024 \\
\textbf{Sodinokibi} & 0.882 & 0.026 & 0.831 & 0.015 & 0.855 & 0.015 \\
\textbf{IcedID} & 0.868 & 0.029 & 0.816 & 0.033 & 0.841 & 0.026 \\
\textbf{QuakBot} & 0.891 & 0.040 & 0.793 & 0.036 & 0.838 & 0.029 \\
\textbf{Gozi} & 0.828 & 0.054 & 0.771 & 0.035 & 0.797 & 0.031 \\
\textbf{AveMariaRAT} & 0.771 & 0.065 & 0.809 & 0.043 & 0.788 & 0.038 \\
\textbf{Pony} & 0.931 & 0.028 & 0.576 & 0.038 & 0.711 & 0.032 \\
\textbf{CobaltStrike} & 0.788 & 0.048 & 0.618 & 0.049 & 0.691 & 0.037 \\
\textbf{Dridex} & 0.526 & 0.032 & 0.938 & 0.015 & 0.673 & 0.026 \\
\textbf{DiamondFox} & 0.597 & 0.049 & 0.737 & 0.037 & 0.658 & 0.033 \\
\textbf{Adware.Generic} & 0.588 & 0.085 & 0.663 & 0.052 & 0.618 & 0.048 \\
\textbf{Amadey} & 0.710 & 0.055 & 0.470 & 0.038 & 0.564 & 0.035 \\
\textbf{OskiStealer} & 0.420 & 0.051 & 0.693 & 0.077 & 0.519 & 0.037 \\
\textbf{ModiLoader} & 0.647 & 0.054 & 0.436 & 0.061 & 0.518 & 0.051 \\
\textbf{njrat} & 0.454 & 0.060 & 0.524 & 0.067 & 0.482 & 0.041 \\
\textbf{ZeuS} & 0.478 & 0.049 & 0.472 & 0.047 & 0.472 & 0.035 \\
\textbf{NetWire} & 0.744 & 0.104 & 0.347 & 0.060 & 0.467 & 0.053 \\
\textbf{BitRAT} & 0.705 & 0.059 & 0.301 & 0.030 & 0.420 & 0.032 \\
\textbf{QuasarRAT} & 0.315 & 0.019 & 0.607 & 0.031 & 0.415 & 0.021 \\
\textbf{Stop} & 0.306 & 0.025 & 0.641 & 0.046 & 0.413 & 0.028 \\
\textbf{AZORult} & 0.789 & 0.093 & 0.280 & 0.039 & 0.412 & 0.048 \\
\textbf{DanaBot} & 0.508 & 0.067 & 0.348 & 0.052 & 0.410 & 0.049 \\
\textbf{SnakeKeylogger} & 0.330 & 0.037 & 0.518 & 0.088 & 0.400 & 0.040 \\
\textbf{FickerStealer} & 0.425 & 0.060 & 0.339 & 0.060 & 0.372 & 0.040 \\
\textbf{AsyncRAT} & 0.442 & 0.069 & 0.315 & 0.044 & 0.366 & 0.044 \\
\textbf{HawkEye} & 0.447 & 0.076 & 0.289 & 0.072 & 0.341 & 0.054 \\
\textbf{MassLogger} & 0.192 & 0.010 & 0.826 & 0.044 & 0.311 & 0.015 \\
\textbf{Smoke Loader} & 0.268 & 0.047 & 0.359 & 0.079 & 0.300 & 0.041 \\
\textbf{ArkeiStealer} & 0.292 & 0.046 & 0.276 & 0.081 & 0.277 & 0.051 \\
\textbf{RaccoonStealer} & 0.271 & 0.053 & 0.251 & 0.068 & 0.256 & 0.052 \\
\textbf{AgentTesla} & 0.311 & 0.083 & 0.179 & 0.065 & 0.224 & 0.071 \\
\textbf{RedLineStealer} & 0.252 & 0.086 & 0.181 & 0.072 & 0.199 & 0.057 \\
\textbf{CryptBot} & 0.263 & 0.084 & 0.150 & 0.062 & 0.182 & 0.055 \\
\textbf{FormBook} & 0.318 & 0.136 & 0.096 & 0.055 & 0.144 & 0.072 \\
\textbf{RemcosRAT} & 0.487 & 0.166 & 0.085 & 0.043 & 0.142 & 0.061 \\
\textbf{Loki} & 0.331 & 0.174 & 0.054 & 0.034 & 0.090 & 0.054 \\
\textbf{NanoCore} & 0.153 & 0.122 & 0.049 & 0.046 & 0.071 & 0.061 \\
\hline
\textbf{Total} & 0.571 & 0.059 & 0.530 & 0.046 & 0.518 & 0.038\\
\hline
\end{tabular}
\caption{D100 - Average outcomes of the multiclass classification and their corresponding standard deviation $\sigma$, displayed according to their F1-score in descending order.}
\label{tab:outcomesmulticlassA}
\end{table}

\begin{figure}[th]
  \includegraphics[width=\linewidth]{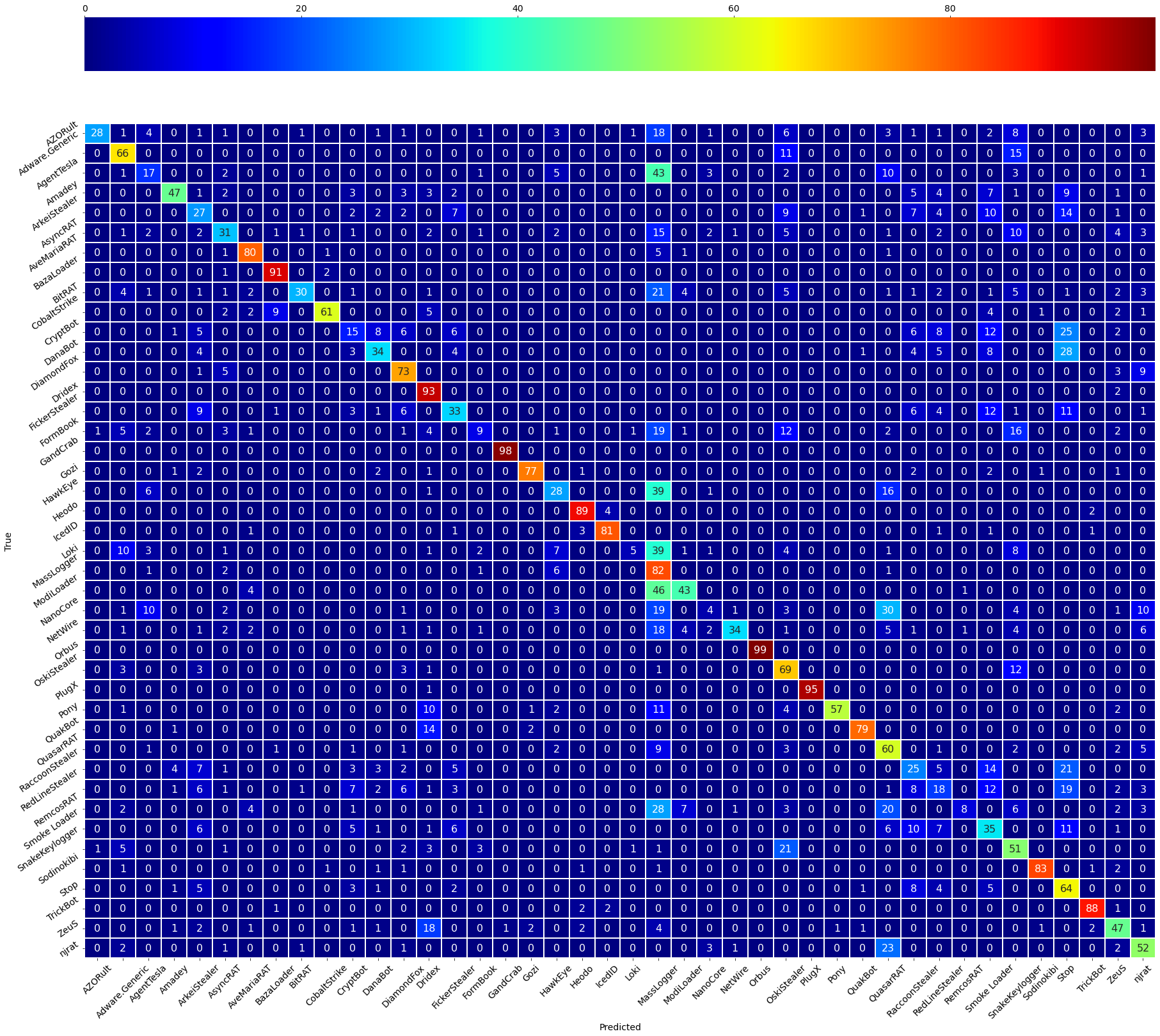}
  \caption{D100 - Confusion matrix of the multiclass classification task.}
  \label{fig:conf_matrixA}
\end{figure}

To provide further insight on the metrics reported by the multiclass classification experiment, we created a confusion matrix with the average classification data and depicted it in Figure \ref{fig:conf_matrixA}. For each row representing a family, the columns show the predicted families by the RF classifier. In a perfect classification, all values in the main diagonal should be equal to the sample size (i.e. 100). Several families were misclassified due to the use of similar features, in some cases overlapping to the extent that made them indistinguishable for the classification method. AgentTesla, AsyncRAT, Formbook, Loki, ModiLoader, NanoCore, RedLineStealer, and RecomcosRat were often misclassified among the families with the most overlapping. We can also observe that in several cases, several malware of the same type (e.g. RATs) are misclassified between them, denoting the possible use of a similar modus operandi. If we observe the table column-wise, many families were misclassified into Adware.Generic, DiamondFox, MassLogger, OskiStealer, QuasarRAT, SmokeLoader, and Stop. Such overlapping showcases the use of the same methods and actions from the malware authors.

To provide a more comprehensive insight about the similarity between families, we used kmeans++ \cite{kmeansplus} on the classification data outcomes and tuned the number of clusters according to the Silhouette metric \cite{rousseeuw1987silhouettes}. By setting the initial number of clusters $k=25$, the malware families were clustered in 15 groups as depicted in Table \ref{tab:clusteringA}. As it can be noted, several families fall into the same groups denoting strong similarities according to the families they are confused with. Isolated families correspond in most cases to the ones that obtained high F1-score in the multiclass classification experiment. Note that different clustering strategies varying the value of $k$ may create slightly different groups. Thus, this experiment aims only to complement the outcomes of the multiclass classification.

\begin{table}[th]
\centering
\rowcolors{2}{gray!25}{white}
\small
\begin{tabular}{c|p{0.7\textwidth}}
\textbf{Group} & \textbf{Families} \\
\hline
1& CryptBot, DanaBot, RaccoonStealer,   RedLineStealer,  Smoke Loader,  Sodinokibi, Stop\\
2& Adware.Generic, AgentTesla, CobaltStrike,  HawkEye,  Loki,  MassLogger,  ModiLoader,  NanoCore,  Pony,   QuasarRAT,   RemcosRAT, SnakeKeylogger\\
3& AZORult, AsyncRAT,  BitRAT,  FormBook,  Heodo,  NetWire,  OskiStealer\\
4& Amadey\\
5& BazaLoader\\
6& AveMariaRAT, QuakBot, TrickBot, njrat\\
7& IcedID\\
8& ZeuS\\
9& Orbus\\
10& Gozi\\
11& PlugX\\
12& DiamondFox\\
13& GandCrab\\
14& Dridex\\
15& ArkeiStealer, FickerStealer\\
 \hline
\end{tabular}
\caption{Kmeans++ clustering applied to D100 with $k=25$. A total of 15 groups were created.}
\label{tab:clusteringA}
\end{table}

\subsubsection{Dataset D300}
\label{exp.setup:mul:b}

In this case, we set up an eligibility threshold of 300 samples to create a new dataset, namely D300. Thus, we selected the 23 families highlighted in purple (see Table \ref{tbl:dataset}). Next, as in the previous experiment, we created a dataset by randomly sampling 300 elements of each family and repeated the experiment 100 times, using 10-fold cross-validation in each experiment. The average outcomes of such an experiment are depicted in Table \ref{tab:outcomesmulticlassB}.

\begin{table}[th]
\centering
\rowcolors{2}{gray!25}{white}
\small
\begin{tabular}{lcc|cc|cc}

\multicolumn{1}{c}{\multirow{1}{*}{\textbf{Family}}} & \multicolumn{2}{c}{\textbf{Precision}} & \multicolumn{2}{c}{\textbf{Recall}} & \multicolumn{2}{c}{\textbf{F1}} \\
\hline
\multicolumn{1}{c}{} & \multicolumn{1}{c}{Average} & \multicolumn{1}{c}{$\sigma$} & \multicolumn{1}{c}{Average} & \multicolumn{1}{c}{$\sigma$} & \multicolumn{1}{c}{Average} & \multicolumn{1}{c}{$\sigma$} \\
\hline
\textbf{Heodo} & 0.928 & 0.012 & 0.966 & 0.010 & 0.947 & 0.008 \\
\textbf{TrickBot} & 0.955 & 0.013 & 0.913 & 0.016 & 0.934 & 0.011 \\
\textbf{QuakBot} & 0.907 & 0.020 & 0.943 & 0.014 & 0.925 & 0.011 \\
\textbf{BazaLoader} & 0.811 & 0.025 & 0.962 & 0.009 & 0.880 & 0.014 \\
\textbf{AveMariaRAT} & 0.808 & 0.035 & 0.852 & 0.017 & 0.829 & 0.019 \\
\textbf{CobaltStrike} & 0.867 & 0.030 & 0.718 & 0.021 & 0.785 & 0.018 \\
\textbf{Gozi} & 0.687 & 0.028 & 0.814 & 0.017 & 0.745 & 0.017 \\
\textbf{ModiLoader} & 0.771 & 0.031 & 0.537 & 0.026 & 0.633 & 0.023 \\
\textbf{njrat} & 0.552 & 0.051 & 0.661 & 0.047 & 0.598 & 0.018 \\
\textbf{NetWire} & 0.740 & 0.078 & 0.381 & 0.026 & 0.501 & 0.020 \\
\textbf{DanaBot} & 0.418 & 0.015 & 0.577 & 0.022 & 0.484 & 0.013 \\
\textbf{ArkeiStealer} & 0.452 & 0.028 & 0.498 & 0.047 & 0.472 & 0.023 \\
\textbf{AZORult} & 0.823 & 0.072 & 0.294 & 0.017 & 0.432 & 0.016 \\
\textbf{MassLogger} & 0.262 & 0.004 & 0.916 & 0.007 & 0.407 & 0.006 \\
\textbf{Smoke Loader} & 0.396 & 0.035 & 0.408 & 0.041 & 0.400 & 0.021 \\
\textbf{RaccoonStealer} & 0.392 & 0.033 & 0.375 & 0.038 & 0.382 & 0.027 \\
\textbf{FormBook} & 0.367 & 0.028 & 0.380 & 0.045 & 0.372 & 0.030 \\
\textbf{RedLineStealer} & 0.365 & 0.065 & 0.262 & 0.054 & 0.298 & 0.033 \\
\textbf{CryptBot} & 0.307 & 0.034 & 0.278 & 0.049 & 0.287 & 0.023 \\
\textbf{AgentTesla} & 0.406 & 0.043 & 0.219 & 0.035 & 0.284 & 0.040 \\
\textbf{NanoCore} & 0.268 & 0.024 & 0.281 & 0.057 & 0.271 & 0.031 \\
\textbf{Loki} & 0.432 & 0.062 & 0.161 & 0.029 & 0.232 & 0.032 \\
\textbf{RemcosRAT} & 0.540 & 0.118 & 0.119 & 0.027 & 0.192 & 0.035 \\
\hline
\textbf{Total} & 0.585 & 0.038 & 0.544 & 0.029 & 0.534 & 0.021\\
\hline
\end{tabular}
\caption{D300 - Average outcomes of the multiclass classification and their corresponding standard deviation $\sigma$, displayed according to their F1-score in descending order.}
\label{tab:outcomesmulticlassB}
\end{table}

\begin{figure}[th]
  \includegraphics[width=\linewidth]{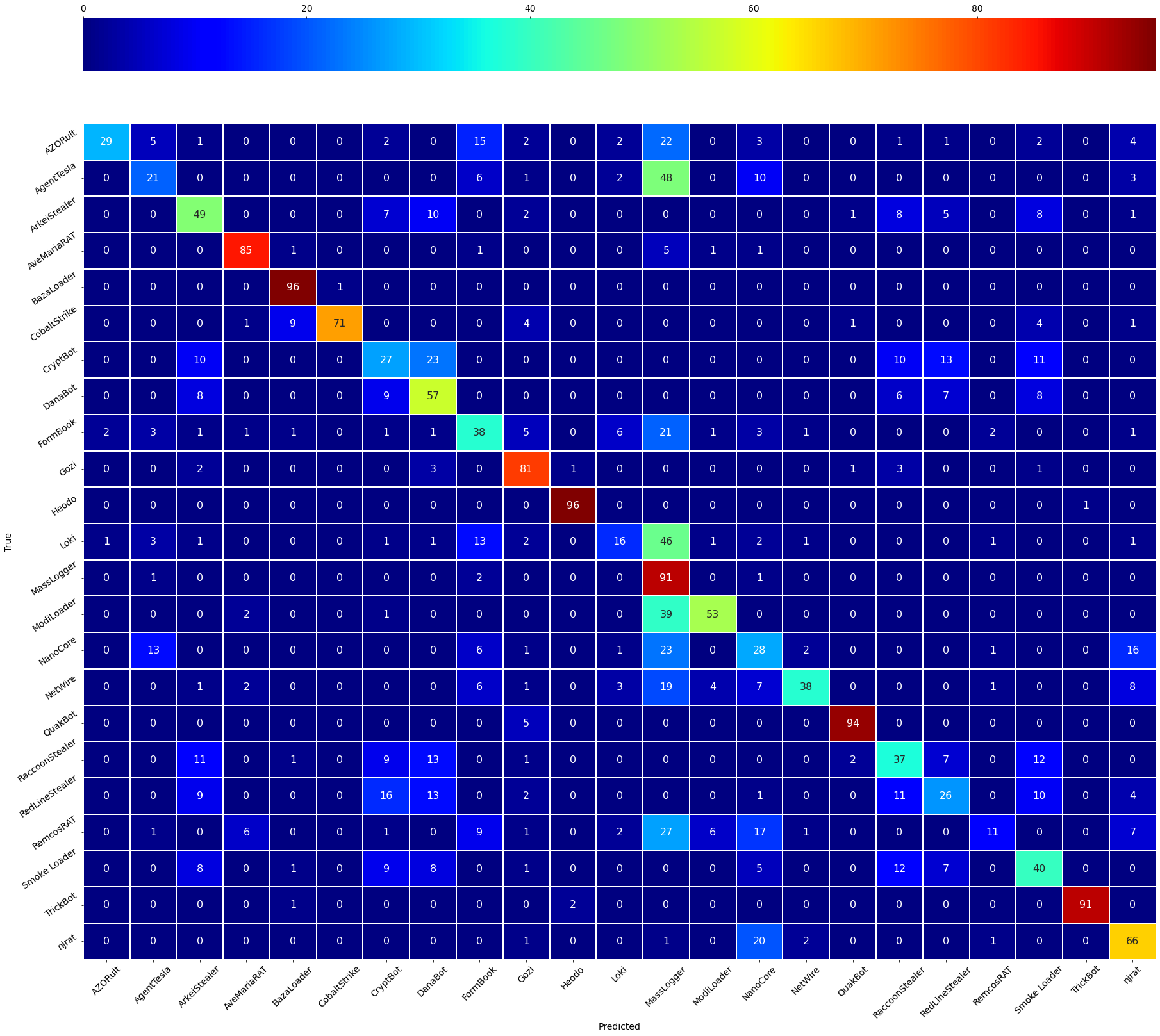}
  \caption{D300 - Confusion matrix of the multiclass classification task.}
  \label{fig:conf_matrixB}
\end{figure}

As it can be observed, the average F1-score is higher than in the case of D100. The latter can be explained due to the use of more samples to leverage the training procedure and the selection of a smaller subset of families, which may reduce the overlapping. As seen in Table \ref{tab:outcomesmulticlassB} and in Figure \ref{fig:conf_matrixB}, most of the families are captured with more reliability than in the case of D100 (see Table \ref{tab:outcomesmulticlassA}). However, since families such as GrandCrab, Orbus, and PlugX, which obtained the best classification outcomes in D100 are not present in D300, the impact on the average values were diminished.

Finally, like in the case of D100, we used kmeans++ to cluster the families of D300 and depicted the outcome in Table \ref{tab:clusteringB}. Notably, the "Stealer" and "Loader" families were clustered together, and we can observe similarities as the ones highlighted in Table \ref{tab:clusteringA}. We used $k=12$ in this experiment since smaller values tend to create groups with high cardinality. As it can be observed, some elements of D100 and D300 are classified in different groups. This is expected behaviour since clustering algorithms will perform differently according to each dataset. Note that the groups created may obey the heuristics of the algorithm, different data distribution, and some elements that would be in a cluster may appear in another according to an algorithm's optimisation metric, e.g. the overall information loss or the inter-cluster similarity \cite{maya2021improving,casino2019privacy}. However, despite such group variations, the distance between the elements will persist since data are the same, and the parameters of clustering algorithms can be tuned to create different grouping strategies according to specific needs.

\begin{table}[th]
\centering
\rowcolors{2}{gray!25}{white}
\small
\begin{tabular}{c|p{0.7\textwidth}}
\textbf{Group} & \textbf{Families} \\
\hline
1& AZORult, AgentTesla, FormBook,   Gozi,  Loki,  NanoCore, NetWire, QuakBot, RemcosRAT\\
2& ArkeiStealer, BazaLoader, CobaltStrike, CryptBot, DanaBot, ModiLoader, RaccoonStealer, RedLineStealer, Smoke Loader\\
3& AveMariaRAT\\
4& njrat\\
5& MassLogger\\
6& Heodo\\
7& TrickBot\\
 \hline
\end{tabular}
\caption{Kmeans++ clustering applied to D300 with $k=12$. A total of 7 groups were created.}
\label{tab:clusteringB}
\end{table}

\section{Discussion}

As discussed in Section \ref{exp.setup:class}, the tuning and training stages are crucial to developing robust models. Note that in the cybersecurity field, obtaining samples and benchmarks can be challenging in several contexts \cite{META}, and authors cannot always have rich and easily obtainable data. In this context, the use of explainable methodologies and a detailed study of the problem that has to be solved are critical to avoid error propagation, which could deliver solutions hindering AI safety \cite{yampolskiy2016artificial}.

While overfitting is a critical issue when tuning the model's hyperparameters, it can be alleviated by using several strategies. First, the fact that the models can obtain accurate classification outcomes even when modifying such parameters denotes that the classes can be differentiated with a certain dynamism. In fact, this was the case since slight variations in the depth's values of RF and XGBoost, and variations in the number of neighbours in the KNN classifier reflected minor changes in the outcomes. However, the outcomes' accuracy decreased dramatically when such parameters were relaxed (for instance, a very low depth). Second, a bias of the models can be detected by using cross-validation and repeating the experiments. As seen in Tables \ref{tab:detailclassification}, \ref{tab:binarysampling}, \ref{tab:outcomesmulticlassA}, and \ref{tab:outcomesmulticlassB} the average standard deviation of the outcomes were very low in most of the cases, denoting stable outcomes.
In cases where the dataset has enough samples to provide robust training and validation, sampling is an effective technique to guarantee unbiased estimates \cite{casino2021intercepting}. We used sampling in both the binary and the multiclass classification several times using random subsets of the original dataset. Thus, we can claim that the reported outcomes represent the dataset distribution in a statistically sound manner.

After analysing the outcomes obtained in Section \ref{exp.setup:class}, we can claim that our method is able to efficiently identify malware samples in a binary classification setup, and identify their family with high reliability in several cases, depending on the number of samples used and the overlapping between similar families. The latter was captured by both our multiclass classification and the clustering experiments, which showcased the possible use of similar strategies and modus operandi leveraged by malicious actors across different families.

In Table \ref{tbl:compareSOA} we compare our approach with the current state of the art in terms of reported metrics, as not all researchers use the same metrics. Note that the proper selection of metrics and the granularity of the reported outcomes is essential to guarantee fair comparison. For instance, when datasets are imbalanced, the accuracy metric may reflect outstanding outcomes if the overrepresented class is classified with high accuracy, even if the other class or classes are not \cite{burnaev2015influence,bekkar2013evaluation}. Clearly, compared to the state of the art, we use the biggest dataset with the most malware families. What is even more interesting in this comparison is that the all these approaches, we had full execution of the binaries, which required execution of at least two minutes and the computational resources of a virtual machine that would resemble a typical host. On the contrary, in our approach, we emulated the execution in dockers managing to scale the experiment smoothly. Moreover, the execution lasted on average almost 10.5sec. Therefore, we managed to have comparable results with state of the art in a heterogeneous and less controlled dataset than our peers in a fraction of the time (on the scale of 1/12) and with better resource allocation.

We opted to provide a comparison based on the methods, not on the dataset; therefore, in one of the state of the art methods the researchers used Android packages and not Windows binaries as the rest. Notably, as we stress in our work, due to the constraints of dynamic analysis in terms of resources, researchers use datasets that are at least one scale smaller than ours. We argue that small datasets with a limited number of families impede methods in real-world cases as the scale factor is crucial. On the contrary, our approach outperforms the current state of the art in all aspects.

\begin{table}[th]
\setlength{\tabcolsep}{3.5pt}
\centering
\scriptsize
\rowcolors{2}{gray!25}{white}
\begin{tabular}{p{0.2in}cclcp{1.7in}}
\toprule
\textbf{Ref.} & \textbf{Scope} & \textbf{Method} & \textbf{Dataset} & \textbf{Classification}& \textbf{Results} \\
\midrule
\cite{Qiao14}& W &C\&M & 3,131M& M(24) & F1 between
0.909 and 0.95 \\
\cite{uppal2014malware}& W&A& 120M/150B &B& Accuracy of 0.985\\
\cite{naval2015employing}&W&E&1,209M/1,316B&B& Accuracy of 0.954\\
 \cite{ki2015novel} & W &D&23,080M/114B& B& Accuracy 0.998\\
 \cite{tang2019dynamic} & W& C & 9,000M &  M(9)&  TPR, precision, recall and
F1 are all $>$99\%, while the FPR is $<$0.1\%\\
 \cite{catak2020deep}&W&C&7107M&  B \& M& F1 score of 0.47 in the multiclass setup, and F1 scores between 0.27 and 0.83 in the binary classification according to different malware types. \\
 \cite{ficco2021malware}&W&C& 4,960M/1,200B& B& Best Accuracies reported between 0.954 and 0.989. \\
 Our work& W & BE & 70,477M/1,059B &B \& M(42)& F1 scores between 0.958 and 0.969 in the binary classification, and between 0.518 and 0.534 in the multiclass setup, in all cases outcomes vary according to the sampling strategy.\\
\bottomrule
\end{tabular}
\caption{Comparison with state of the art.\\
\textbf{Notation:} Scope: (A)ndroid, (W)indows. Method: A:https://www.apimonitor.com/ C: Cuckoo, D: Detours \cite{271603} M: Maleur, E: Ether \cite{dinaburg2008ether} BE: Binary Emunlation. Dataset: We report the number of (M)alicious and (B)enign samples that were used. Classification: (B)inary, (M)ulticlass and (\#families). The latter classifies samples into worms, trojans, droppers, etc.}
\label{tbl:compareSOA}
\end{table}
\subsection{Comparison with a commercial sandbox}
 Intrigued by the exceptional results of the binary analysis in a real-world dataset, we opted to compare our approach to a commercial sandbox. This way, beyond showing the efficacy of our approach, we also overcome several limitations and biases of performing such a task on our own. For instance, such a task requires a lot of resources for a significant amount of time. Moreover, configuring the sandbox properly to detect and bypass modern malware evasion and anti-analysis methods is not trivial and subject to a lot of hardware, software, and configuration constraints.\\
Given that MalwareBazaar\footnote{\url{https://bazaar.abuse.ch/about/}} has good interaction with  Hatching's Triage\footnote{\url{https://tria.ge/}}; the collected samples are submitted for analysis in their platform, we opted to collect the sandbox analysis results from it and compare its verdict with ours. Each sample for Windows in Triage is executed in two VMs, one with Windows 7 and one with Windows 10. Moreover, the platform performs some static tests and returns a score on the scale of one to ten of how malicious it was detected. As a rule of thumb, scores below five are not considered malicious. Therefore, we opted to collect the scores for each sample in our dataset from the API that Triage provides and use it to assess our efficacy. At this point, we have to note that since Triage provides at least two scores for each sample, one may request later to analyse the sample again; we decided to record the highest and minimum score for each sample. Nonetheless, even if it is not in our favour, we keep the highest score for the comparison.\\
Given that the analysis of samples from MalwareBazaar was not integrated from the very beginning of the platform, 8740 samples from our dataset were not analysed from Triage. Moreover, since the benign files that we use in our dataset are already shipped with Windows, their hashes are known and included in databases such as the National Software Reference Library (NSRL)\footnote{\url{https://www.nist.gov/itl/ssd/software-quality-group/national-software-reference-library-nsrl}} and xCyclopedia\footnote{\url{https://strontic.github.io/xcyclopedia/}}. Practically, they would immediately be reported as benign as they are also signed by Microsoft; therefore, we believe that they should not be scanned from a sandbox. The scores of the remaining 61701 malicious samples are reported in Table \ref{tbl:scores_triage}. Practically, 5306
 (8.6\% of the files analysed in Triage) managed to bypass the sandbox analysis and were classified as benign. On the contrary, approximately 0.1\% of the malicious samples were misclassified in our analysis, as seen in the granular information depicted in Table 3 (i.e. we report the outcomes per class; thus, malicious classification values reported a recall value of 0.999. Moreover, since Triage devotes 2.5 minutes per VM by default, the user may prolong the execution time per VM; at least 5 minutes are used per sample. Without comparing the resources, which for a sandbox are far greater as they imply the dedicated use of a VM, the sandbox approach of Triage is 28.57 times slower than ours.
\begin{table}[th]
    \centering
    \begin{tabular}{lrr}
    \toprule
    \textbf{Score}&\textbf{Samples$_M$}&\textbf{Samples$_m$}\\
    \midrule
1&4531&7030\\
2 &  0& 0\\
3&549  & 810\\
4 &226 & 156\\
5&245  & 323\\
6&1640 & 1773\\
7&966  & 1291\\
8&3126 & 3560\\
9&310  &202 \\
10&50108 &46556\\\bottomrule
\textbf{Total} & &\textbf{61701}\\\bottomrule
    \end{tabular}
    \caption{Distribution of scores of the samples from Triage. Samples$_M$ denotes the number of samples with the highest score specified in the first column. Similarly, Samples$_m$ denotes the number of samples with lowest score specified in the first column.}
    \label{tbl:scores_triage}
\end{table}
\subsection{Limitations}
While we argue that binary emulation is an excellent complimentary solution for traditional sandbox analysis, it has several limitations and constraints. Firstly, due to its maturity, one cannot analyse arbitrarily any file nor for every operating system. Indeed, in our experiments, we opted to use only a part of the EXE files. More precisely, we kept only the Windows PE files, non-dynamic libraries, unmanaged code executables. This choice is because unmanaged code executables compile straight to machine code and are directly executed by the operating system, so binary emulation is more straightforward. Using other formats, e.g. Windows PE files based on .NET framework, would create issues for the binary emulation at the current state. Thus, this analysis approach cannot cater to the needs of arbitrary binary emulation at this stage. Nonetheless, as the frameworks mature, we expect these gaps to be filled.
We also had to skip Dynamic-link libraries because they cannot be executed directly. For a PE file to be executed, an entry point (main function) is required for the execution. In addition to the above, as many community-powered frameworks, Qiling has not implemented all Windows APIs, Linux, and Darwin syscalls yet. As a result, API-complex programs are failing to complete an end-to-end binary emulation. The Qiling community continuously implements and submits new API that we expect will result in finer-grained results, more detailed results, and even new capabilities.

\section{Conclusions}
With the continuous advances in malware design and deployment, early detection and automated analysis of malware become an imminent need. The latter part is rather a time and resource-consuming task as, for instance, executing a binary in a sandbox environment requires the replication of an actual user host in a virtual machine, the hooking of all calls and the monitoring of all its interactions in the network, file system level, API and system calls etc. As a result, a significant amount of resources have to be dedicated to the execution of a single binary for a considerable amount of time. When the amount of daily malware samples is on the scale of 450,000, such approaches cannot meet this rate. The latter has to be considered in parallel to the fact that static analysis can be easily bypassed. As a result, we need to find ways to significantly boost the speed of dynamic analysis, or at least make it scale more efficiently.

One of the approaches that have been fostered the past few years is binary emulation. Practically, the execution of a binary is emulated, so there is no need to utilise a whole virtual machine. Additionally, the analyst may interrupt the execution of the binary and manipulate it to explore additional execution paths. To the best of our knowledge, this work performs the first evaluation of binary emulation in malware classification. Our automated pipeline allows us to extract the necessary features from the emulated execution to perform both binary and multiclass classification. Our extensive experiments with a large real-world dataset illustrate that this line of research is rather promising. Our machine learning-based approach manages to achieve outstanding results in binary classification and to provide rather accurate results for many malware families in the case of multiclass classification. Indeed, for binary classification, our approach outperforms a commercial sandbox. Notably, we achieve the above with only a fraction of the resources and time that sandboxing solutions would require.
Notably, this is verified with a real-world experiment against a commercial sandbox. We argue that the above proves the prevalence of our method compared to the state of the art and practice.

Due to the current maturity of existing binary emulation frameworks, not all syscalls have been implemented, so the collected information is a fragment of their actual potential. We argue that we managed to have such outstanding results with binary emulation in such early stages in a real-world dataset with diverse malware families against a mature and commercial sandbox clearly indicates that the results are robust and subject to further greater improvements. The study of different feature selection methods and their performance in the context of binary emulation in malware classification is an interesting research line that we will explore in future work. Since this efficacy is coupled with a significant reduction in resource allocation at a fraction of the time of sandbox approaches, it implies that this line of research can efficiently complement current approaches to timely counter malware spread. Following this line of research, in the future, we plan to explore the exploitation of further static features along with the dynamic ones in the context of binary emulation. Moreover, we plan to explore the effectiveness of symbolic and concolic execution in scale to determine whether they can meet the requirements of real-world scenarios.

\section*{Acknowledgements}
This work was supported by the European Commission under the Horizon 2020 Programme (H2020), as part of the projects  \textit{LOCARD} (Grant Agreement no. 832735), HEROES (Grant Agreement no. 101021801), and CyberSec4Europe (Grant Agreement no. 830929). Fran Casino was supported by the Beatriu de Pinós programme of the Government of Catalonia (Grant No. 2020 BP 00035).\\

The content of this article does not reflect the official opinion of the European Union. Responsibility for the information and views expressed therein lies entirely with the authors.

\end{document}